\documentstyle[preprint,aps,draft,psfig]{revtex}

\newcommand{\ket}[1]{\left | \, #1 \right \rangle}
\newcommand{\bra}[1]{\left \langle #1 \, \right |}

\def\one{\leavevmode\hbox{\normalsize1\kern-4.6pt\large1}}

\begin{document}
\draft

\title{Quantum privacy amplification and the security of quantum
  cryptography over noisy channels}

\author{David Deutsch $^{(1)}$, Artur Ekert $^{(1)}$, Richard Jozsa $^{(2)}$,\\
Chiara Macchiavello $^{(1)}$, Sandu Popescu $^{(3)}$, Anna Sanpera $^{(1)}$}

\address{$(1)$ Clarendon Laboratory, Department of Physics, University
  of Oxford, Parks Road, Oxford OX1 3PU, UK}
\address{$(2)$ School of Mathematics and Statistics, University of
  Plymouth, Plymouth, Devon PL4 8AA, UK}
\address{$(3)$ Department of Electrical, Computer and Systems
  Engineering, Boston University, Boston MA 02215, U.S.A.}

\date{Received \today}

\maketitle

\begin{abstract} 
\noindent
  Existing quantum cryptographic schemes are not, as they stand,
  operable in the presence of noise on the quantum communication
  channel. Although they become operable if they are supplemented by
  classical privacy-amplification techniques, the resulting schemes
  are difficult to analyse and have not been proved secure. We
  introduce the concept of quantum privacy amplification and a
  cryptographic scheme incorporating it which is provably secure over
  a noisy channel.  The scheme uses an `entanglement purification'
  procedure which, because it requires only a few quantum
  Controlled-Not and single-qubit operations, could be implemented
  using technology that is currently being developed. The scheme
  allows an arbitrarily small bound to be placed on the information
  that any eavesdropper may extract from the encrypted message.
\end{abstract}


Quantum cryptography~\cite{BB84,E91,B92} allows two parties
(traditionally known as Alice and Bob) to establish a secure random
cryptographic key if, firstly, they have access to a quantum
communication channel, and secondly, they can exchange classical
public messages which can be monitored but not altered by an
eavesdropper (Eve). Using such a key, a secure message of equal length
can be transmitted over the classical channel. However, the security
of quantum cryptography has so far been proved only for the idealised
case where the quantum channel, in the absence of eavesdropping, is
{\em noiseless}. That is because, under existing protocols, Alice and
Bob detect eavesdropping by performing certain quantum measurements on
transmitted batches of qubits and then using statistical tests to
determine, with any desired degree of confidence, that the transmitted
qubits are not entangled with any third system such as Eve. The
problem is that there is in principle no way of distinguishing
entanglement with an eavesdropper (caused by her measurements) from
entanglement with the environment caused by innocent {\em noise}, some
of which is presumably always present.

This implies that all existing protocols are, strictly speaking,
inoperable in the presence of noise, since they require the
transmission of messages to be suspended whenever an eavesdropper (or,
therefore, noise) is detected. Conversely, if we want a protocol that
is secure in the presence of noise, we must find one that allows
secure transmission to continue even in the presence of eavesdroppers.
To this end, one might consider modifying the existing protocols by
reducing the statistical confidence level at which Alice and Bob
accept a batch of qubits. Instead of the astronomically high level
envisaged in the idealised protocol, they would set the level so that
they would accept most batches that had encountered a given level of
noise. They would then have to assume that some of the information in
the batch was known to an eavesdropper. It seems reasonable that
classical privacy amplification \cite{classical_privacy_amplification}
could then be used to distil, from large numbers of such qubits, a key
in whose security one could have an astronomically high level of
confidence. However, no such scheme has yet been proved to be secure.
Existing proofs of the security of classical privacy amplification
apply only to classical communication channels and classical
eavesdroppers. They do not cover the new eavesdropping strategies that
become possible in the quantum case: for instance, causing a quantum
ancilla to interact with the encrypted message, storing the ancilla
and later performing a measurement on it that is chosen according to
the data that Alice and Bob exchange publicly.

In this paper we present a protocol that is secure in the presence of
noise and an eavesdropper. It uses entanglement-based quantum
cryptography~\cite{E91}, but with a new element, an `entanglement
purification' procedure. This allows Alice and Bob to generate a pair
of qubits in a state that is close to a pure, maximally entangled
state, and whose entanglement with any outside system is arbitrarily
low. They can generate this from any supply of pairs of qubits in
mixed states with non-zero entanglement, even if an eavesdropper has
had access to those qubits.

Our procedure -- a {\it Quantum Privacy Amplification} algorithm --
can be performed by Alice and Bob at distant locations by a sequence
of local operations which are agreed upon by communication over a
public channel. It is related to the procedure described in
\cite{pur}, but is much more efficient.

In the idealised theory of entanglement-based quantum cryptography,
Alice and Bob have a supply of qubit-pairs, each pair being in the
pure, maximally entangled state $\ket{\phi^+}$, where

\begin{eqnarray}
\left.
{\ket{\phi^\pm}  =\frac {1}{\sqrt{2}} (\ket{00}\pm\ket{11})}
\atop
{\ket{\psi^\pm} =\frac {1}{\sqrt{2}} (\ket{01}\pm\ket{10})}
\right\}
\label{bell_basis}
\end{eqnarray}

These are the so-called `Bell states' which form a convenient basis
for the state space of a qubit-pair. Alice and Bob each have one qubit
from each pair. In the presence of noise, each pair would in general
have become entangled with other pairs and with the environment, and
would be described by a density operator on the space spanned by
(\ref{bell_basis}).

Note that any two qubits that are jointly in a pure state cannot be
entangled with any third physical object.  Therefore any algorithm
that delivers qubit-pairs in pure states must also have eliminated the
entanglement between any of those pairs and any other system. Our
scheme is based on an iterative quantum algorithm which, if performed
with perfect accuracy, starting with a collection of qubit-pairs in
mixed states, would discard some of them and leave the remaining ones
in states converging to $\ket{\phi^+}\bra{\phi^+}$. If (as must be the
case realistically) the algorithm is performed imperfectly, the
density operator of the pairs remaining after each iteration will not
converge on $\ket{\phi^+}\bra{\phi^+}$, but will fluctuate in a
neighbourhood of it. However, we shall argue that the degree of
entanglement with any eavesdropper may nevertheless continue to fall,
and can be brought to an arbitrary low value even though the
purification to $\ket{\phi^+}\bra{\phi^+}$ remains imperfect.

Our first departure from existing quantum cryptographic schemes is to
assume that Eve {\em does} interact with all the qubits that are
transmitted or received by either Alice or Bob. Indeed we analyse the
scenario that is most favourable for eavesdropping, namely where Eve
herself is allowed to prepare all the qubit pairs that Alice and Bob
will subsequently use for cryptography. Any realistic situation would
also involve environmental noise that is not under Eve's control, but
this may be treated as a special case in which Eve is not using the
full information available to her.

Suppose, then, that Eve has prepared two qubit pairs in some manner of
her own choosing, and sends one qubit from each pair to each of Alice
and Bob. 
Let the density operators of the two pairs be ${\hat
\rho}$ and
${\hat \rho}'$ respectively. Alice performs a unitary operation
\begin{eqnarray}
\ket{0} &\longrightarrow & \frac{1}{\sqrt 2} (\ket{0} -i\ket{1})\\
\ket{1} &\longrightarrow & \frac{1}{\sqrt 2} (\ket{1} -i\ket{0})
\end{eqnarray}
on each of her two qubits; Bob performs the inverse operation
\begin{eqnarray}
\ket{0} &\longrightarrow & \frac{1}{\sqrt 2} (\ket{0} +i\ket{1})\\
\ket{1} &\longrightarrow & \frac{1}{\sqrt 2} (\ket{1} +i\ket{0})
\end{eqnarray}
on his. If the qubits are spin-$\frac{1}{2}$
particles and the computation basis is that of the eigenstates of the
$z$~components of their spins, then the two operations correspond respectively
to rotations by $\pi/2$ and
$-\pi/2$ about the
$x$ axis.

Then Alice and Bob each perform two instances of the quantum
Controlled-Not operation
\begin{equation}
\stackrel{\mbox{\tiny{control}}}{\ket{a}}\,\,\,
\stackrel{\mbox{\tiny{target}}}{\ket{b}}
\longrightarrow
\stackrel{\mbox{\tiny{control}}}{\ket{a}}
\stackrel{\mbox{\tiny{target}}}{\ket{a\oplus
b}}
\qquad (a,b)\in
\{0,1\}
\end{equation}

where one pair (${\hat \rho}$) comprises the two control qubits and
the other one (${\hat \rho}'$) the two target qubits~\cite{BDEJ}.
Alice and Bob then measure the target qubits in the computational
basis (e.g. they measure the $z$~components of the targets' spins). If
the outcomes coincide (e.g. both spins up or both spins down) they
keep the control pair for the next round, and discard the target pair.
If the outcomes do not coincide, both pairs are discarded.

To see the effect of this procedure, consider the case in which each
pair is in state ${\hat\rho}$ (although the joint state of the two
pairs need not be the simple product ${\hat\rho} \otimes {\hat\rho }$
as they may be entangled with each other). This case will suffice for
our applications.  We express the density operator ${\hat \rho}$ in
the Bell basis
$\{\ket{\phi^+},\ket{\psi^-},\ket{\psi^+},\ket{\phi^-}\}$ and denote
by $\{A,B,C,D\}$ the diagonal elements in that basis. Note that the
first diagonal element $A= \bra{\phi^+ } {\hat \rho} \ket{\phi^+ }$,
which we call the `fidelity', is the probability that the qubit would
pass a test for being in the state $\ket{\phi^+ }$.  Thus we wish to
drive the fidelity to 1 (which implies that the other three diagonal
elements go to 0). Now, in the case where the control qubits are
retained, their density operator ${\hat \rho}\:\tilde{}$, will have
diagonal elements $\{\tilde A,\tilde B,\tilde C,\tilde D\}$ which
depend on average {\em only} on the diagonal elements of ${\hat
  \rho}$:
\begin{equation} 
\left.
\begin{array}{ccc}
{\tilde A} & = & \frac{A^2 +B^2}{N}\\
{\tilde B} & = & \frac{2CD}{N}\\
{\tilde C} & = & \frac{C^2 +D^2}{N}\\
{\tilde D} & = & \frac{2AB}{N},
\end{array}
\right\}
\label{m4}
\end{equation}
where $N=(A+B)^2 +(C+D)^2$ is the probability 
that Alice and Bob obtain coinciding outcomes in the measurements on
the target pair.
That is, if the procedure is carried out many times on an ensemble of
such pairs of pairs, then $\tilde{A},\tilde{B},\tilde{C}$ and
$\tilde{D}$ give the average diagonal entries of the surviving pairs.
Note that if the average $\tilde{A}$ is driven to 1 then each of the
surviving pairs must individually approach the pure state
$\ket{\phi^+}\bra{\phi^+}$.

In passing we note that if the two input pairs have {\em different} states
${\hat\rho}$ and ${\hat\rho'}$ with diagonal elements $\{ A,B,C,D\}$
and $\{A',B',C',D'\}$ respectively, then the retained control pairs
will, on average, have diagonal elements given by:
\begin{equation} 
\left.
\begin{array}{ccc}
{\tilde A} & = & \frac{AA'+BB'}{N}\\
{\tilde B} & = & \frac{C'D+CD'}{N}\\
{\tilde C} & = & \frac{CC'+DD'}{N}\\
{\tilde D} & = & \frac{AB'+A'B}{N},
\end{array}
\right\}
\label{mmmm4}
\end{equation}
where $N=(A+B)(A'+B')+(C+D)(C'+D')$, which generalises (\ref{m4}).

Suppose that Eve has provided $L$ pairs of qubits, with density
operators ${\hat\rho}_1$, ${\hat \rho}_2, ..., {\hat \rho}_{L}$. (This
is {\em not} to say that their overall density operator is ${\hat
  \rho}_1 \otimes {\hat \rho}_2 \otimes ... \otimes {\hat \rho}_{L}$,
for Eve may have prepared them in an entangled state.)  Alice and Bob
know nothing about the state preparation, they are simply presented
with an ensemble of $L$ pairs of qubits from which they can (if they
wish) estimate the average density operator ${\hat\rho}_{ave}$:
\begin{equation}
{\hat \rho}_{ave} = \frac{1}{L} \left({\hat \rho}_1 + {\hat \rho}_2 +
... +{\hat \rho}_{L}\right).
\end{equation}
which characterises the ensemble of pairs.

Alice and Bob now select pairs at random from the ensemble of provided
pairs and apply the QPA procedure to pairs of these selected pairs.
Thus we may set ${\hat\rho} = {\hat\rho}_{ave}$ in (\ref{m4}) and we
are in effect studying the properties of the map
\begin{equation}
\left (
\begin{array}{c}
A\\B\\C\\D
\end{array}
\right ) 
\longrightarrow
\left (
\begin{array}{c}
\tilde A\\
\tilde B\\
\tilde C\\
\tilde D
\end{array}
\right )=
\frac{1}{N}  
\left (
\begin{array}{c}
A^2 + B^2\\
2CD\\
C^2+D^2\\
2AB
\end{array}
\right )
\label{av4}
\end{equation}

on the average diagonal elements of density operators (in the Bell
basis). $(\tilde{A},\tilde{B},\tilde{C},\tilde{D})$ in (\ref{av4})
gives the average diagonal entries for the states of the surviving
pairs i.e. the diagonal entries of the average density operator of the
ensemble of surviving pairs. Therefore the repeated application of the
QPA procedure -- generating successive ensembles of surviving pairs --
corresponds to iteration of the map in (\ref{av4}).

Several interesting properties of this map can be easily verified.
For example if at any stage the fidelity $A$ exceeds $\frac{1}{2}$,
then after one more iteration, it still exceeds $\frac{1}{2}$.
Although $A$ does not necessarily increase monotonically, our target
point, $A=1$, $B=C=D=0$, is a fixed point of the map, and is the only
fixed point in the region $A>\frac{1}{2}$. It is a local attractor. We
have been unable to obtain a proof that it is also a global attractor
in the region $A>\frac{1}{2}$, but we have verified this by computer
simulation. In other words, if we begin with pairs whose average
fidelity exceeds $\frac{1}{2}$, but which are otherwise in an
arbitrary state containing arbitrary correlations with each other and
with an eavesdropper, then the states of pairs surviving after
successive iterations always converge to the unit-fidelity pure state
$\ket{\phi^+}$. Since this is a pure state, none of the surviving
pairs is, in the limit, entangled with any other system.

\begin{figure}
\centerline{
\psfig{width=186mm,file=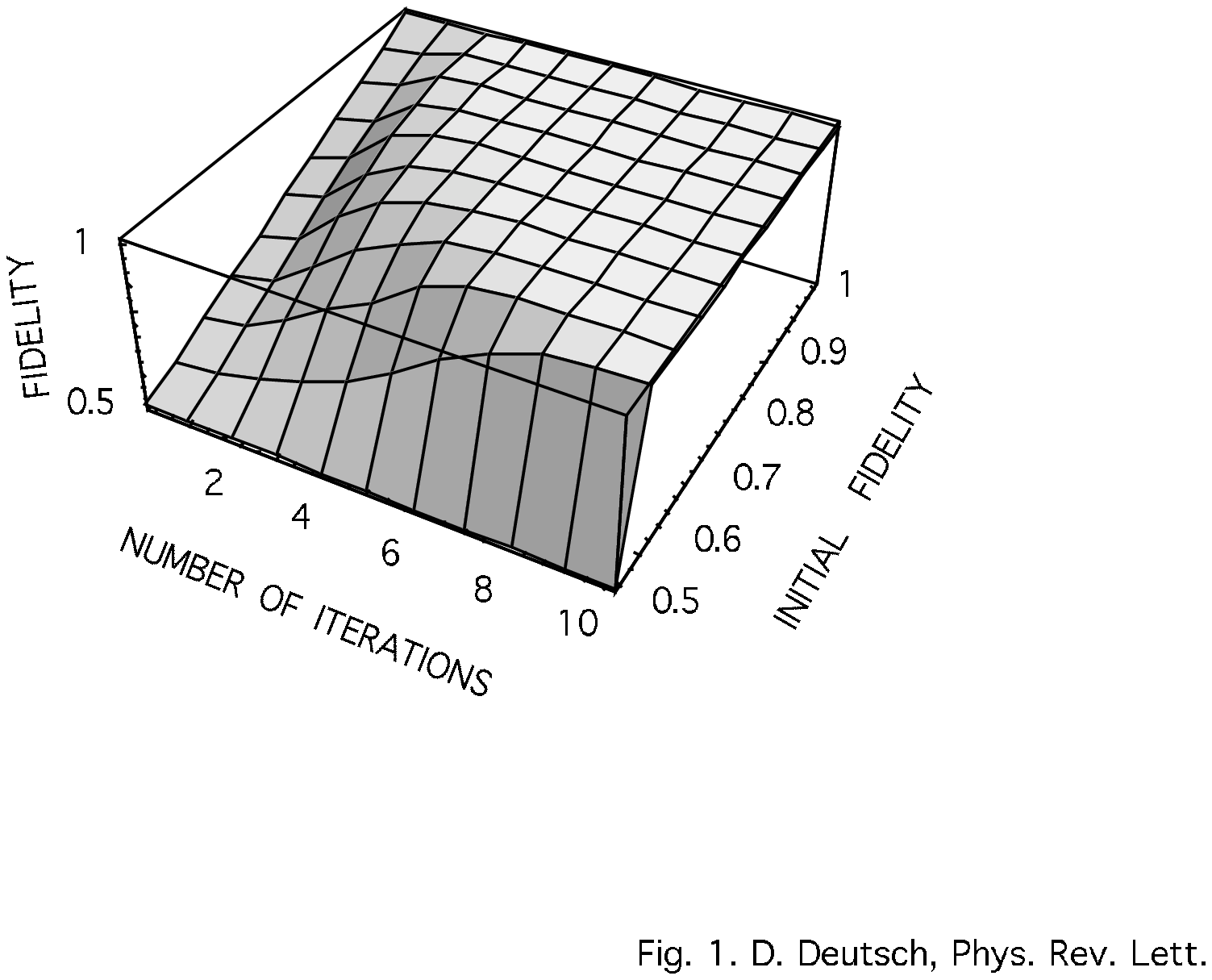}
}
\vspace{2mm}
\caption[f1]{Average fidelity as a function of the initial fidelity and
  the number of iterations.}
\label{f:3d}
\end{figure}

To illustrate the behaviour of the iteration, in Fig.(\ref{f:3d}) we
plot the fidelity as a function of the initial fidelity and the number
of iterations, in cases where $A>\frac{1}{2}$ and $B=C=D$ initially.

If the procedure were performed only imperfectly, then as we have
said, the fidelity would approach some value below $1$, and would then
fluctuate.  However this does not necessarily imply an associated
level of residual entanglement with the eavesdropper.  Let us consider
more closely what it means to perform the QPA procedure `imperfectly'.
We may at least assume that Alice and Bob are capable of performing
{\em local} computations in secret, and therefore that even an
imperfect QPA apparatus {\em does not interact with Eve}. In other
words, the perturbing interactions that make each QPA step imperfect
are local to Alice or Bob's private domains. (The issue of the
security of these private domains is beyond the remit of cryptology.)
Consider a class of perturbations, which may include both
imperfections of the measurements and of the quantum logic gates, for
which the net effect is as if all the QPA steps were performed
perfectly, but some local interactions took place before and after the
steps were performed. Such interactions would reduce the fidelity of
the surviving qubit pairs, but could not increase their entanglement
with the eavesdropper.  Indeed they could not prevent the elimination
of such entanglement in successive QPA steps.  In this scenario, even
though the purification will be limited by the accuracy of the logic
gates and detectors, the {\em entanglement with the eavesdropper}, on
which her opportunity to read the key entirely depends, nevertheless
becomes arbitrarily small.  Specifically, if the procedure is
performed with moderate accuracy, then her information due to
entanglement must fall roughly as if the accuracy were perfect.

\begin{figure}
\centerline{
\psfig{width=156mm,file=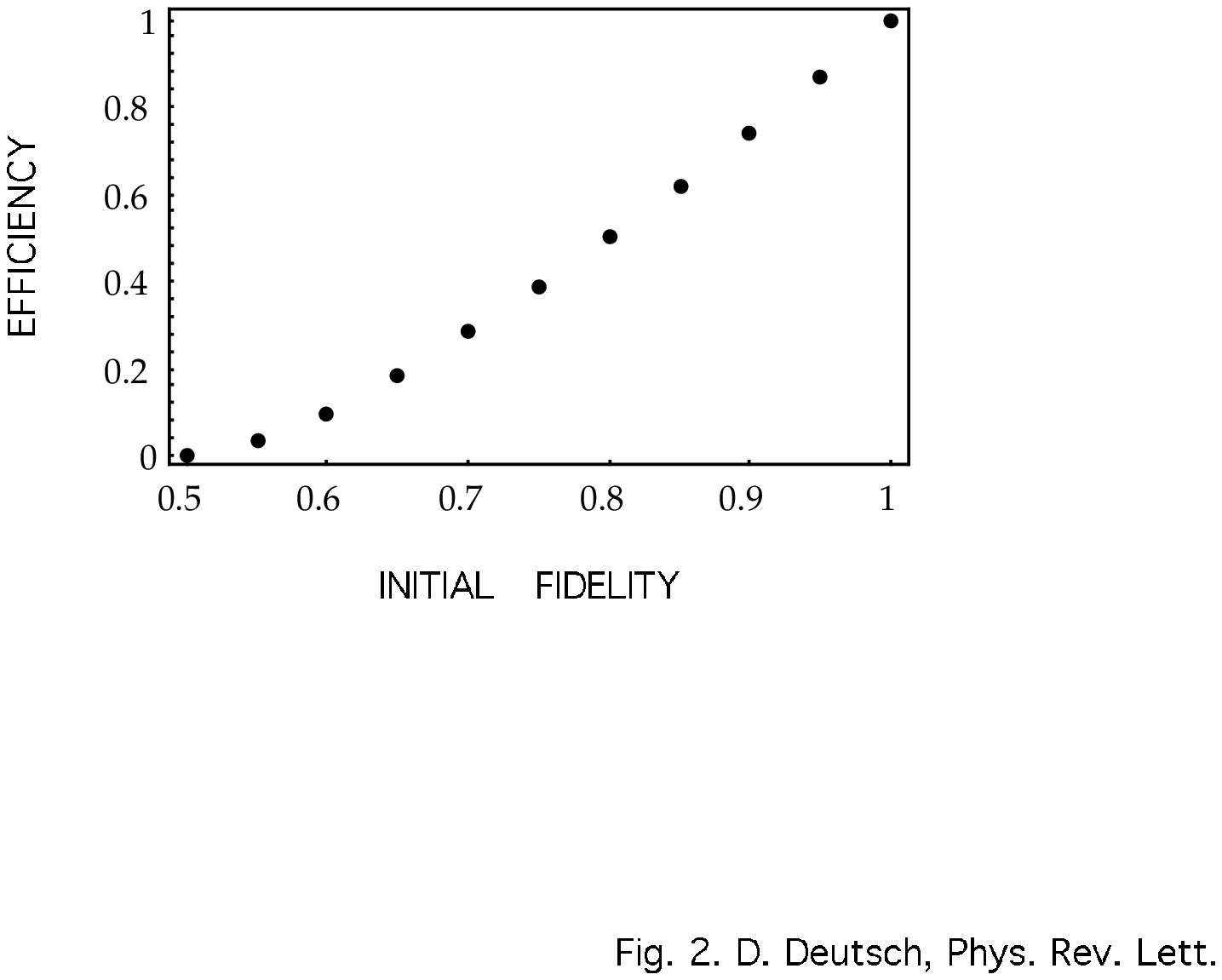}
}
\caption[f2]{Proportion of purified pairs left by the QPA algorithm as a function
  of the initial fidelity in units of $2^{-10}$.}
\label{f:eff}
\end{figure}

The QPA procedure is rather wasteful in terms of discarded particles -
at least one half of the particles (the ones used as controls) is lost
at every iteration.  In Fig.~\ref{f:eff} we plot the efficiency, i.e.
the proportion of the initial supply of pairs that remain, after 10
iterations, in units of $2^{-10}$, as a function of the initial
fidelity for initial states with $B=C=D$. Still the efficiency of our
scheme compares favourably with the entanglement purification scheme
as described in~\cite{pur} (about 1000 times more efficient for $A$
close to $0.5$) and it can be directly applied to purify states which
are not necessarily of the Werner form~\cite{Werner}.

The QPA is capable of purifying (or disentangling) a collection of
pairs in any state ${\hat \rho}$ whose average fidelity with respect
to at least one maximally entangled state (i.e. a Bell state or a
state obtained from a Bell state via local unitary operations) is
greater than $\frac{1}{2}$ (because any state of that type can be
transformed into $\ket{\phi^+}$ via local unitary
operations~\cite{Wiesner-Bennett}). If we denote by $\cal{B}$ a class
of pure, maximally entangled states (the generalised Bell states) then
the condition that the state $\hat\rho$ can be purified using the QPA
is

\begin{equation}
\max_{\phi\in\cal{B}} \bra{\phi}{\hat \rho} \ket{\phi}>{1\over 2}.
\label{cond}
\end{equation}
N.B., this condition is not equivalent to the Horodecki
condition~\cite{HHH} characterising mixed states which can violate a
generalised Bell inequality (CHSH inequality~\cite{chsh}). Indeed
there exist mixed states which satisfy {\em both} our condition
(\ref{cond}) {\em and} the CHSH inequalities. Thus, analysis of the
QPA reveals a more complete characterisation of non-locality than that
given by Bell's theorem (c.f. also~\cite{Pop1,Pop2,Gis}). We hope to
elaborate this in a forthcoming paper.

The practical implementation of the QPA would require efficient
quantum Controlled-Not gates operating directly on information
carriers. Perhaps the most promising implementation of gates of this
type (in the QPA context) is the one proposed by Turchette {\em et
  al.}~\cite{turch}. It operates on polarised photons and allows the
polarisation of the target photon to be rotated depending on the
polarisation of the control photon. Although the current efficiency of
the device is quite low, recent experimental progress in this field
raises hopes for a successful QPA experiment in the not too distant
future.

This research was supported in part by Elsag-Bailey plc. We would like
to thank A.~Barenco and W.K.~Wootters for stimulating discussions.
A.E. and R.J.  are sponsored by The Royal Society, London.  C.M. is
sponsored by the European Union HCM Programme.  A. S. is sponsored by
the Fleming Foundation.  A.E., R.J. and S.P acknowledge Rabezzana
Grignolino d'Asti.

\end{document}